\documentclass[a4paper,oneside,12pt]{article}
\usepackage[cp1250]{inputenc}
\usepackage[T1]{fontenc}
\usepackage{amsmath}
\usepackage[dvips]{graphicx}
\usepackage{algorithmic}
\usepackage{algorithm}
\usepackage{hyperref}
\usepackage{url}
\usepackage{calc}
\newlength{\mgrlewy}\newlength{\mgrprawy}
\newlength{\mgrgora}\newlength{\mgrdol}
\newlength{\mgrstopka}\newlength{\mgrdoglowki}
\newlength{\pomc}
\newcount\tempcntr
 \advance\pomc-.5\baselineskip
 \divide\pomc\baselineskip
 \tempcntr=\pomc
\begin{document}
\title{Toy model for large non - symmetric random matrices}
\author{Małgorzata Snarska$^1$\thanks{\url{snarska@th.if.uj.edu.pl}}}
\date{\today}
\maketitle
{\small{$ $\\$^1$Mark Kac Complex Systems Research Center.Institute of Physics, Jagellonian
University, \\ 30-059 Cracow, Reymonta 4, Poland}
\begin{abstract}
\noindent
Non-symmetric rectangular correlation matrices occur in many problems in economics.
We test the method of extracting statistically meaningful correlations between input and
output variables of large dimensionality and build a toy model for artificially included
correlations in large random time series.The results are then applied to analysis of polish
macroeconomic data and can be used as an alternative to classical cointegration approach.
\end{abstract}
\vspace{0.5cm}
\fussy
\section{Multivariate modeling of time series - setting the stage}
Multivariate time series data are widely available in different fields like economics, finance, medicine or telecommunication.
Building efficient multivariate models,which help understanding the relation between a large number of possible causes and resulting effects,
is therefore crucial for many decision - making activities.\\
Due to the works of Granger \cite{Granger}, Bollerslev \cite{Bollerslev} and Sims \cite{Sims}
Vector Autoregressive (VAR) and Vector GARCH (eg.BEKK,VEC) models are nowadays deeply investigated especially in the field of econometrics.
It is believed that, the system itself should determine the number of relevant input and output factors.The \emph{" brute force"} method involves taking all the possible input and output factors and systematically correlate them, hoping to find some signal. One can easily convince oneself, that VAR and Vector GARCH models work well for small number of input and output variables, however suffer from the so called \emph{"dimensionality curse"} i.e. blow up with just a few factors. The cross - equation correlation matrix contains all the information about contemporaneous correlation in a Vector model and may be its greatest strength and its greatest asset. Since no questionable a priori assumptions are imposed , fitting a Vector model allows dataset to speak for itself. Still without imposing any restrictions on the structure of the correlation matrix one cannot make a causal interpretation of the results.
We believe there exist highly non-trivial statistically meaningful correlations
between two samples of non-equal size (i.e. input and output
variables of large dimensionality), which can be then treated as "natural" restrictions for the correlations matrix structure. Since however the data inside the samples can also be correlated, one has to remove in-the-sample correlations first and then find some signal (if any) outside the samples.
\section{Model description}
The detailed description of the ideas, that drive our toy model can be found in \cite{Bouchaud}.
The authors suggested to compare the singular value spectrum of the empirical rectangular $M\times N$ correlation matrix with a benchmark obtained using Random Matrix Theory results (c.f. \cite{Edelman}), assuming there are no correlation between the variables.
\subsection{Notation and mathematical aspects}
Consider $N$ input factors $X_a$ $a=1,\dots, N$ and $M$ output factors $Y_{\alpha}$ $\alpha =1,\ldots,M$ with the total number of observations being $T$. All time series are standardized to have zero mean and unit variance. The data can be completely different or be the same variables but observed at different times. \\To remove the correlations inside each sample we form two correlation matrices,which contain information about in-the-sample correlations.
\begin{equation}\label{internal}
\mathbf{C_X}=\frac{1}{T}X^TX \qquad \mathbf{C_Y}=\frac{1}{T}Y^TY
\end{equation}
The matrices are then diagonalized and the empirical spectrum is compared to the theoretical Mar\v{c}enko-Pastur spectrum \cite{Marcenko},\cite{Laloux},\cite{Burda}. This allows to find and extract statistically significant factors.The eigenvalues,which lie much below the lower edge of the Mar\v{c}enko-Pastur spectrum represent the redundant factors, rejected by the system. They can be excluded from further analysis. which slightly reduces the dimensionality of the problem (i.e. one gets rid of spurious correlations).However before doing that, one has to create a set of uncorrelated unit variance input variables $\hat{X}$ and output variables  $\hat{Y}$.
\begin{equation}\label{result}
\hat{X}_{at}=\frac{1}{\sqrt{T\lambda_a}}V^TX_t\qquad \hat{Y}_{\alpha t}=\frac{1}{\sqrt{T\lambda_{\alpha}}}U^TY_t
\end{equation}
where $V,U$,  $\lambda_a$, $\lambda_{\alpha}$ are the corresponding eigenvectors and eigenvalues of $C_X$ , $C_Y$ respectively.
\\ Now we are ready to create the $M\times N$ cross-correlation matrix $G$ between the $\hat{Y}$ and $\hat{X}$
\begin{equation}
G=\hat{Y}\hat{X}^T
\end{equation}
which includes only the correlations between input and output factors. The singular value decomposition (SVD) (c.f. \cite{Friedberg}) is used to find the empirical spectrum of eigenvalues.The singular value
spectrum represent the strength of cross-correlations between input and output factors.
\subsection{Singular values and Free Random Matrix Theory}
Theoretical predictions for eigenvalue density are obtained using the Free Random Matrix Theory and assuming no correlations between the samples. The final result for singular eigenvalue density, when there are no correlations between input and output data is:
\begin{equation}
\varrho(s) =\max \left(1-\frac{N}{T}, 1-\frac{M}{T}\right)\delta(s) + \max \left(\frac{M+N-T}{T},0 \right)\delta(s-1)+Re\frac{\sqrt{(s^2-\gamma_-)(\gamma_+-s^2)}}{\pi s(1-s^2)}
\end{equation}
with $s$ being singular eigenvalues and
\[\gamma_{\pm}=\frac{1}{T^2}\left(NT+MT - 2MN \pm 2\sqrt{MN(T-N)(T-M)}\right)\]
Empirical results are then compared with the above  benchmark. Any exceptions may suggest nontrivial correlations between the samples.
\section{Applications}
Two different set of data were investigated: Polish macroeconomic data and generated set of data, where temporal cross - correlations are introduced by construction.
\subsection{Polish Macroeconomic data}
The analysis began with checking, whether the method described in \cite{Bouchaud} is relevant for describing the relation between the inflation indexes for polish macroeconomic indexes and other polish macroeconomic data published by different government and non-government agencies. We have used monthly $M=13$ changes of different CPI indicators as our predicted variables (i.e. output sample $Y$) and $N=48$ monthly changes of economic indicators (eg. sectoral employment, foreign exchange reserves, PPI's) as explanatory variables. The investigated period was between 01.1999-08.2007 (i.e. $T=104$).The data were standardized, but the factors in input and output samples were not selected very carefully, so the data could speak for themselves and system could be able to select the optimal combination of variables. The next step involved cleaning internal correlations in each sample. To do it, we have used equation (\ref{internal}).The resulting matrices were then diagonalized and two sets of internally uncorrelated data were prepared.
\begin{figure}[h]
\begin{tabular}{cc}
  \includegraphics[width=7cm]{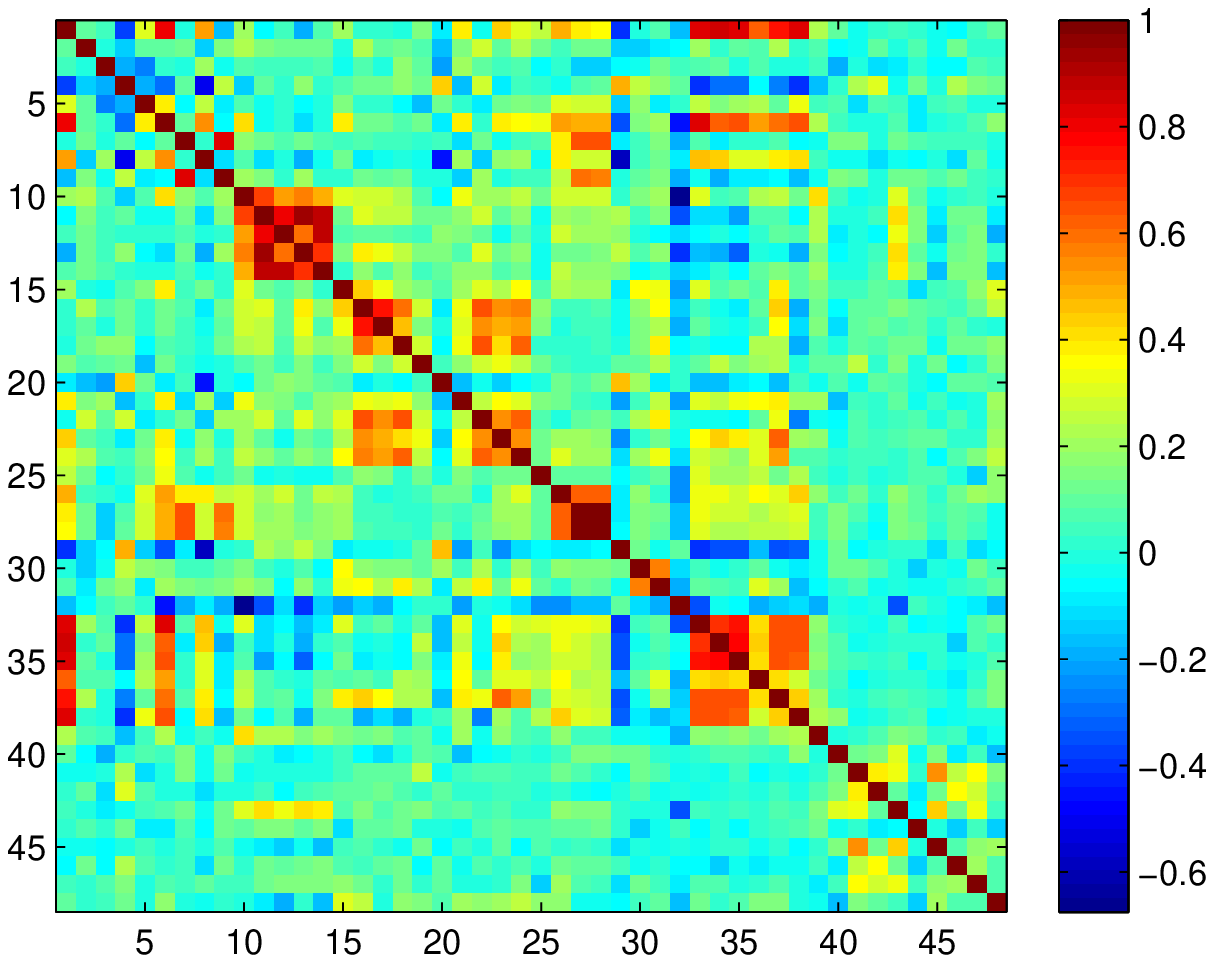} &\includegraphics [width=7cm]{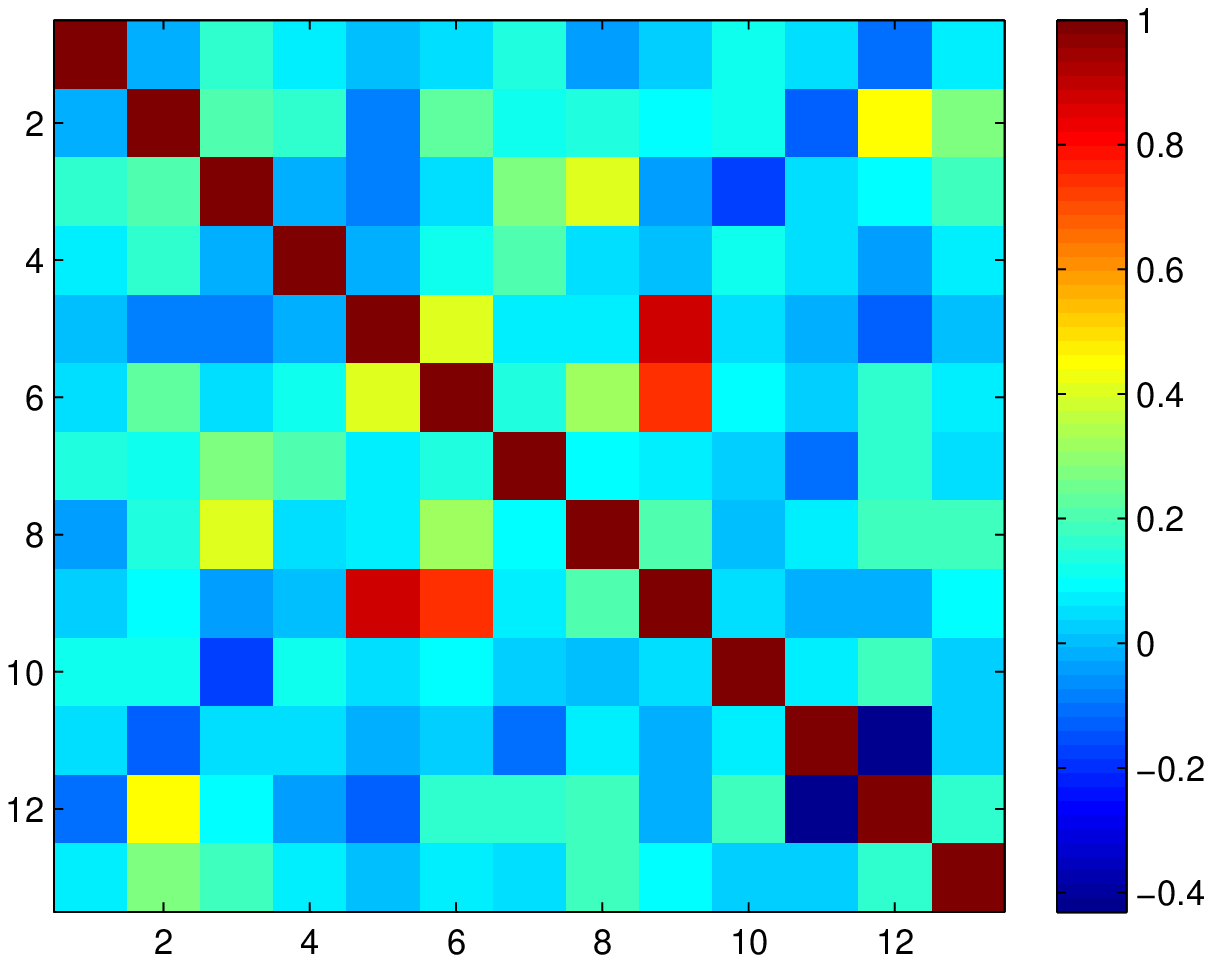}\\
  \small{48 macroeconomic indicators - input $X$} & \small{13 CPI's - output $Y$}\\
\end{tabular}
\caption{Correlation matrices representing in-the-sample correlations}
\end{figure}
From the uncorrelated data we create the rectangular matrix $G$ and diagonalize it to calculate singular eigenvalues. Finally we have used the benchmark calculated in equation (\ref{result}) to compare the data with the predicted eigenvalue density.The results show, that there exists some singular eigenvalues, which do not fit the benchmark. Among them, the highest singular eigenvalue $s_1=2.5$ and the corresponding singular eigenvector, represent standard correlation between expenses for electricity and producers prices in the energy sector. There are however other non-trivial relations between eg. CPI in telecommunication sector and foreign exchange reserves.
 \begin{figure}[h]
\begin{tabular}{cc}
  \includegraphics[width=7cm]{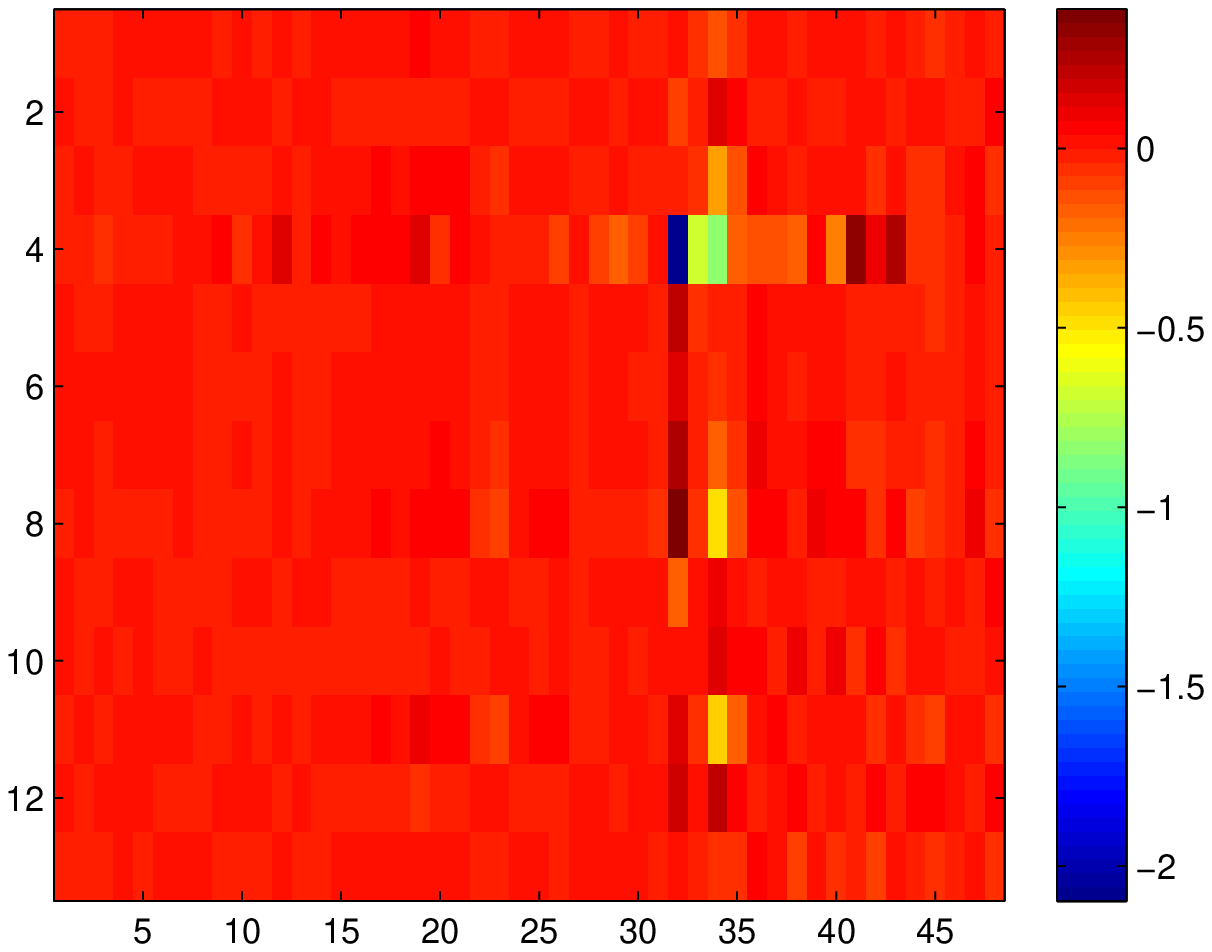} &\includegraphics [width=7cm]{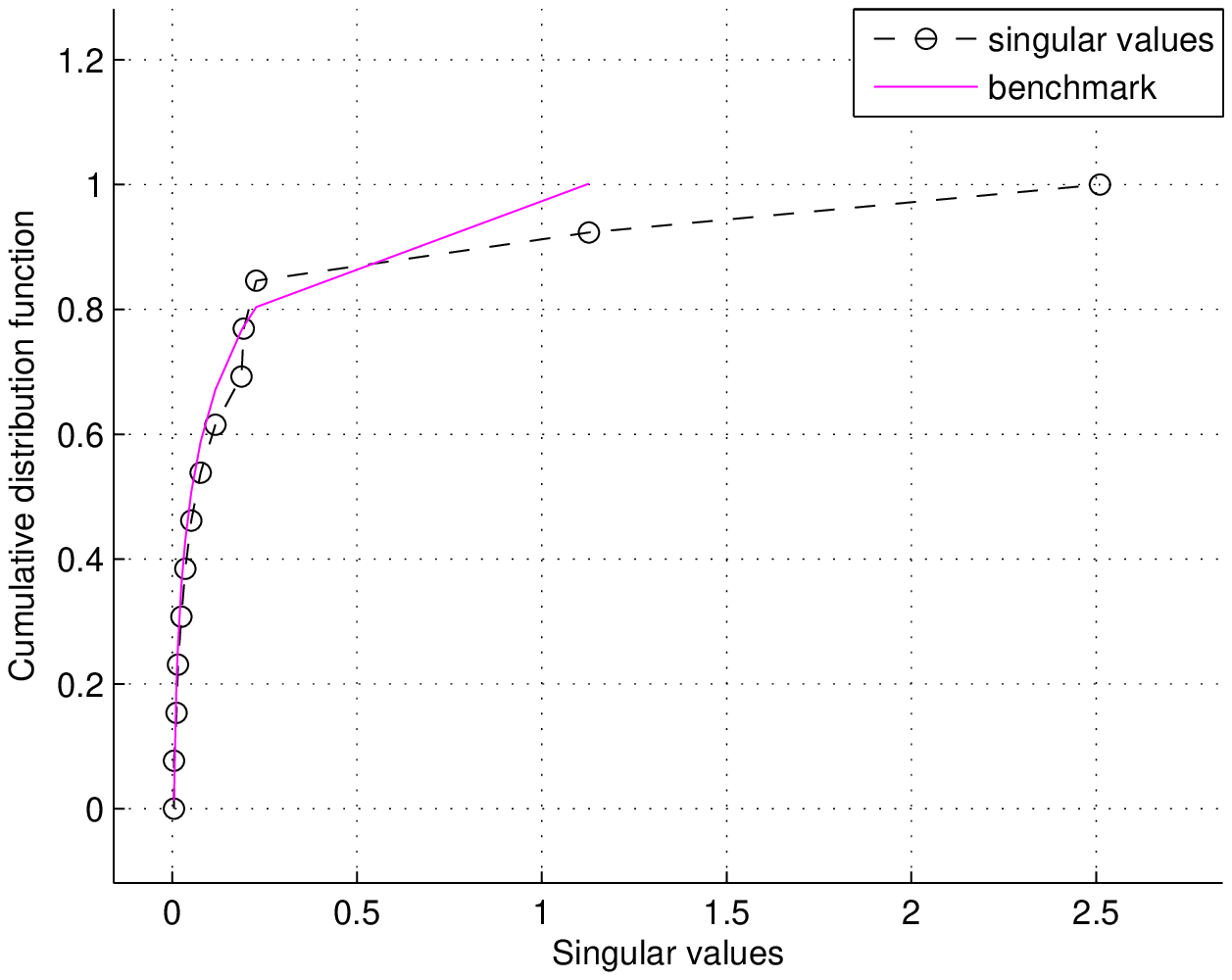}\\
  \small{Matrix $G$ - out of the sample correlations} & \small{Singular eigenvalues of $G$}\\
\end{tabular}
\caption{Out of the sample correlations}
\end{figure}
\subsection{Artificially generated data - multivariate GARCH (1,1) process}
We also wanted to check whether the above method was able to extract temporal correlations for the data, that memorize its past realizations and, but not necessary, its past variances. In order to proceed a sample of 100 paths of GARCH(1,1) type and 1000 observations were generated. The steps presented in the previous section were repeated. The input data were 100 GARCH(1,1) paths lagged by one observation, and the output data were represented by the original set of variables.
As a result we got one eigenvalue, which do not fit well the assumed benchmark and is suspected to represent the memory of the data. However further test to confirm the idea are still necessary.
 \begin{figure}[h]
 \begin{center}
  \includegraphics[width=10cm]{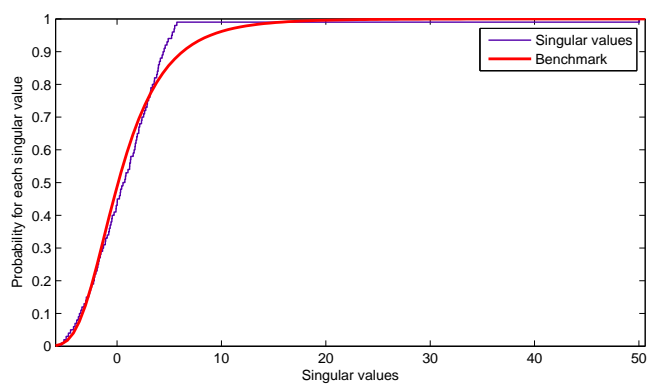}
\caption{\small{Singular eigenvalues from the GARCH process compared to the benchmark}}
\end{center}
\end{figure}
\section*{Conclusions and Future Work}
\addcontentsline{toc}{chapter}{Conclusions and Future Work}
Both examples show that there exists non - trivial correlation structure between input and output variables.Though redundant factors add significant amount of noise in the problem, the SVD decomposition allows to find only truly informative factors. This might be helpful in analyzing the effect of so called sunspot or spurious correlations and investigation of correlations between different stock exchanges,  and will be the part of our future work.

\end{document}